\documentclass[12pt]{elsarticle}
\usepackage[utf8]{inputenc}
\usepackage{graphicx}
\usepackage{subfigure}

\usepackage{amssymb}
\usepackage{amsthm}
\usepackage{url}
\usepackage{amsmath}
\usepackage{subfigure}
\usepackage{textcomp}
\usepackage{xcolor}

\graphicspath{{figs/}}

\journal{BioSystems}

\begin{document}

\begin{frontmatter}

\title{On electrical gates on fungal colony}
\author[1,2]{Alexander E. Beasley}
\author[3]{Phil Ayres}
\author[4]{Martin Tegelaar}
\author[2]{Michail-Antisthenis Tsompanas}
\author[2]{Andrew Adamatzky}

\address[1]{Centre for Engineering Research, University of Hertfordshire, UK}
\address[2]{Unconventional Computing Laboratory, UWE, Bristol, UK}
\address[3]{The Centre for Information Technology and Architecture, Royal Danish Academy, Copenhagen, Denmark}
\address[4]{Microbiology Department, University of Utrecht, Utrecht, The Netherlands}

\begin{abstract}
\noindent
Mycelium networks are promising substrates for designing unconventional computing devices providing rich topologies and geometries where signals propagate and interact. Fulfilling our long-term objectives of prototyping electrical analog computers from living mycelium networks, including networks hybridised with nanoparticles, we explore the possibility of implementing Boolean logical gates based on electrical properties of fungal colonies. We converted a 3D image-data stack of \emph{Aspergillus  niger} fungal colony to an Euclidean graph and modelled the colony as resistive and capacitive (RC) networks, where electrical parameters of edges were functions of the edges' lengths. We found that {\sc and}, {\sc or} and {\sc and-not} gates are implementable in RC networks derived from the geometrical structure of the real fungal colony. 
\end{abstract}


\end{frontmatter}

\section{Introduction}

Fungi are demonstrated to be at the forefront of environmentally sustainable biomaterials~\cite{karana2018material,jones2020engineered,cerimi2019fungi} used in manufacturing of acoustic~\cite{pelletier2013evaluation,elsacker2020comprehensive,robertson2020fungal} and thermal~\cite{yang2017physical,xing2018growing,girometta2019physico,dias2021investigation,wang2016experimental,cardenas2020thermal} insulation panels, packaging materials~\cite{holt2012fungal,sivaprasad2021development,mojumdar2021mushroom} and adaptive wearables~\cite{adamatzky2021reactive,silverman2020development,karana2018material,appels2020use,jones2020leather}. In our project `Fungal architectures'~\cite{adamatzky2019fungal} we proposed to grow mycelium bound composites into monolithic building elements~\cite{adamatzky5adaptive}. The composite would combine living mycelium, capable of sensing light, chemicals, gases, gravity and electric fields~\cite{bahn2007sensing,van2002arbuscular,kung2005possible,fomina2000negative,bahn2006co2,jaffe2002thigmo,howitz2008xenohormesis}, with dead mycelium functionalised using nanoparticles and polymers. These living building structures would have embedded bioelectronics electronics~\cite{beasley2020capacitive,beasley2020mem,beasley2020fungal}, implement sensorial fusion and decision making in the mycelium networks~\cite{adamatzky2020boolean} and be able to grow monolithic buildings from the functionalised fungal substrate~\cite{adamatzky5adaptive}. 

A decision making feature requires inference logical circuits to be embedded directly into mycelium bonded composites. To check what range and frequencies of logical gates could be implemented in the mycelium bound composites we adopted an approach developed originally in \cite{adamatzky2020boolean,siccardi2020actin}. The technique is based on selecting a pair of input sites, applying all possible combinations of inputs to the sites and recording outputs on a set of the selected output sites. 
The approach belongs to same family of computation outsourcing techniques as  \emph{in materio} computing~\cite{miller2002evolution,miller2014evolution,stepney2019co,miller2018materio,miller2019alchemy} and reservoir computing~\cite{verstraeten2007experimental,lukovsevivcius2009reservoir,dale2017reservoir,konkoli2018reservoir,dale2019substrate}. In our previous studies \cite{adamatzky2020boolean} we demonstrated that logical circuits can be derived from electrical spiking activity of the fungal colony. The approach, whilst elegant theoretically, might lack practical applications because the spiking activity of living fungi is of very low frequency, e.g. a spike per 20 minutes~\cite{adamatzky2018spiking,adamatzky2021electrical}. Thus, we decided to explore electrical properties of the fungal colony, because the electrical analog implementation of logical gates is notoriously fast.  In the numerical experiments described here, `0' and `1' signals are represented by low and high voltage applied to the input sites.

\section{Methods}
\label{methods}

\begin{figure}[!tbp]
    \centering
    \includegraphics[width=0.98\textwidth]{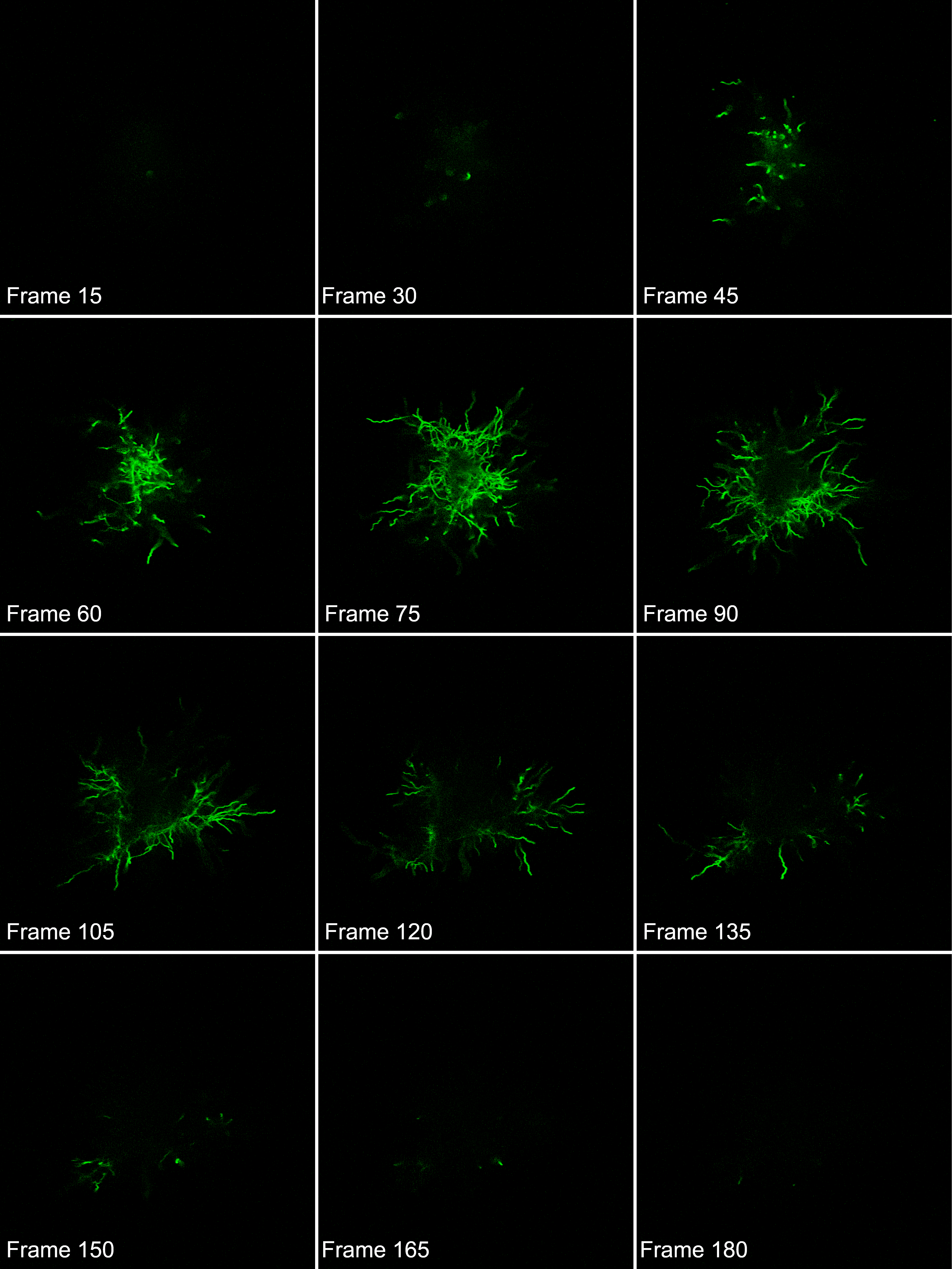}
    \caption{Z-slices of the fungal colony of \emph{Aspergillus niger} imaged by fluorescence microscopy.}
    \label{fig:colony}
\end{figure}

\begin{figure}[!tbp]
    \centering
    \includegraphics[width=0.96\textwidth]{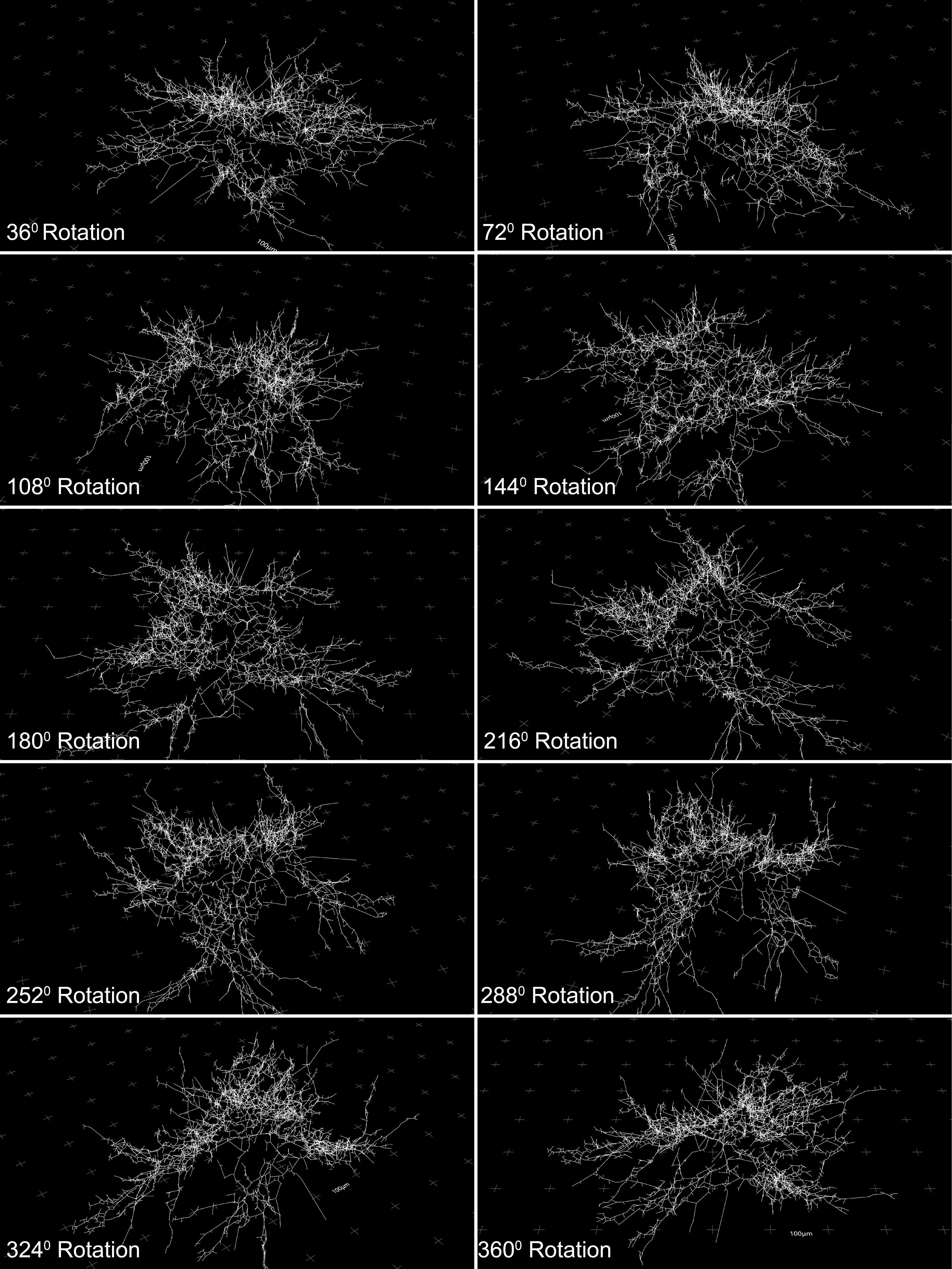}
    \caption{Perspective views of the 3D Graph. Each frame shows the graph after a 36$^{\circ}$ rotation around the z-axis with origin located approximately in the centre of the colony, on the x-y plane indicated with registration marks.}
    \label{fig:graph}
\end{figure}

A 3D colony  of \emph{Aspergillus niger} fungus was cultured with fluorescent protein and visualised by confocal microscopy as detailed in Sect.~\ref{appendix}. Z-stacks of imaged micro-colonies were provided using 100 slices with a slice thickness of 8.35~µm (Fig.~\ref{fig:colony}). The Z-stacks of the colony have been converted to a 3D graph (Fig.~\ref{fig:graph}) as detailed in Sect.~\ref{appendix}. The 3D graph was converted to an resistive and capacitive (RC) network, whose magnitudes are a function of the length of the connections. Resistances were in the order of kOhms and capacitance were in the order of pF. Separate models were created with the RC connections modelled either in series or in parallel modules. The networks were parsed for the order one nodes, which are considered to the extent of the sample. The positive voltage  and ground nodes were randomly assigned from the sample and 1000 networks are created in each arrangement for analysis. SPICE analysis consisted of transient analysis using a two voltage pulses of 60~mV on the randomly assigned positive nodes with the following parameters: $T\textsubscript{delay}$= 10~s for $V_1$ and 20~s for $V_2$, $T\textsubscript{rise}$ = 0.001~s, $T\textsubscript{fall}$ = 0.001~s, $T\textsubscript{on}$ = 10~s for $V_1$ and  20~s for $V_2$, $T\textsubscript{off}$ = 20~s for $V_1$ and 20~s for $V_2$, $N\textsubscript{cycles}$ = 2 for $V_1$ and 1 for $V_2$. 
Circuit analysis was transient analysis for 40~s in steps of 1~ms. The voltage at each node and current through each link were measured every 1~ms of the simulation. We modelled the fungal colony in serial RC networks and parallel RC networks.

\section{Results}
\label{results}

\begin{figure}[!tbp]
    \centering
    \subfigure[]{\includegraphics[width=0.99\textwidth]{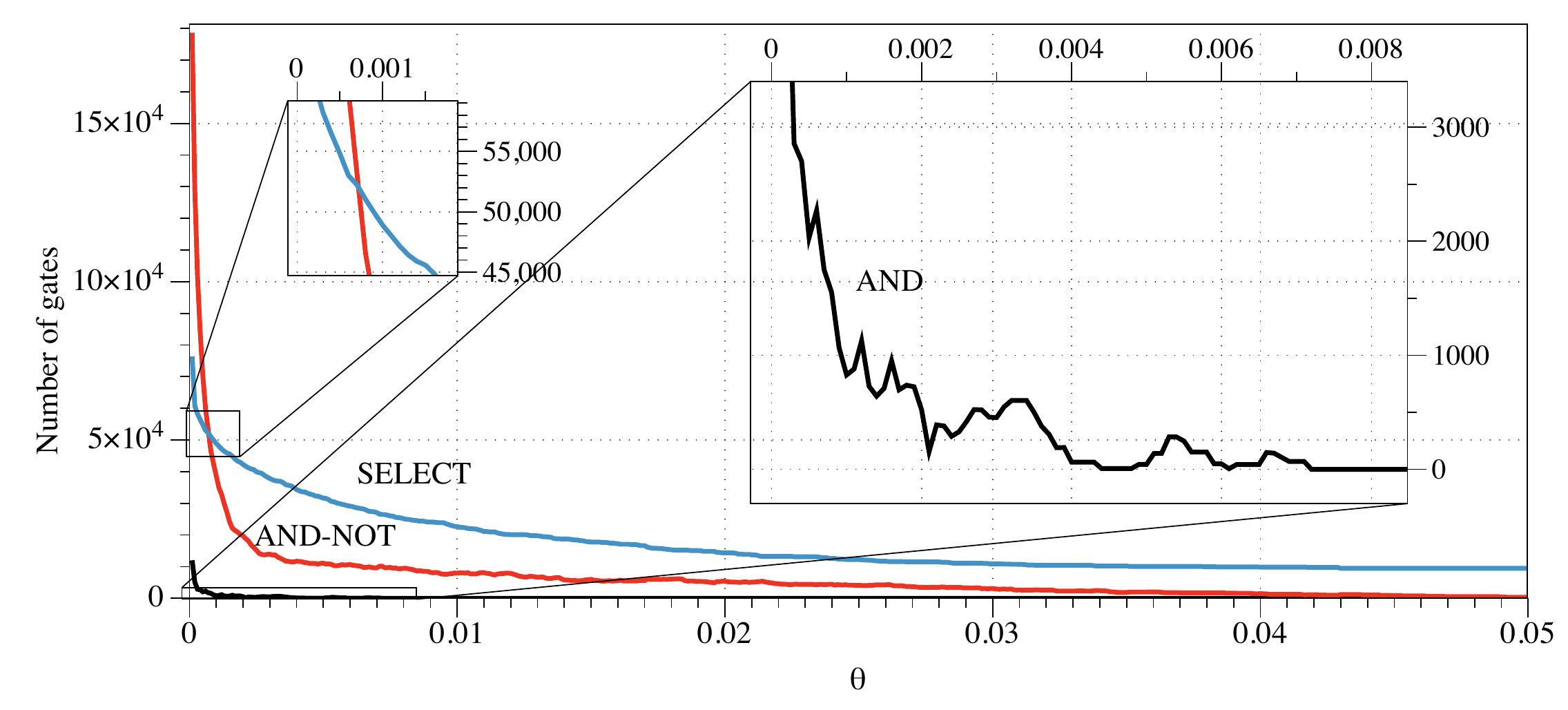}}
    \subfigure[]{\includegraphics[width=0.99\textwidth]{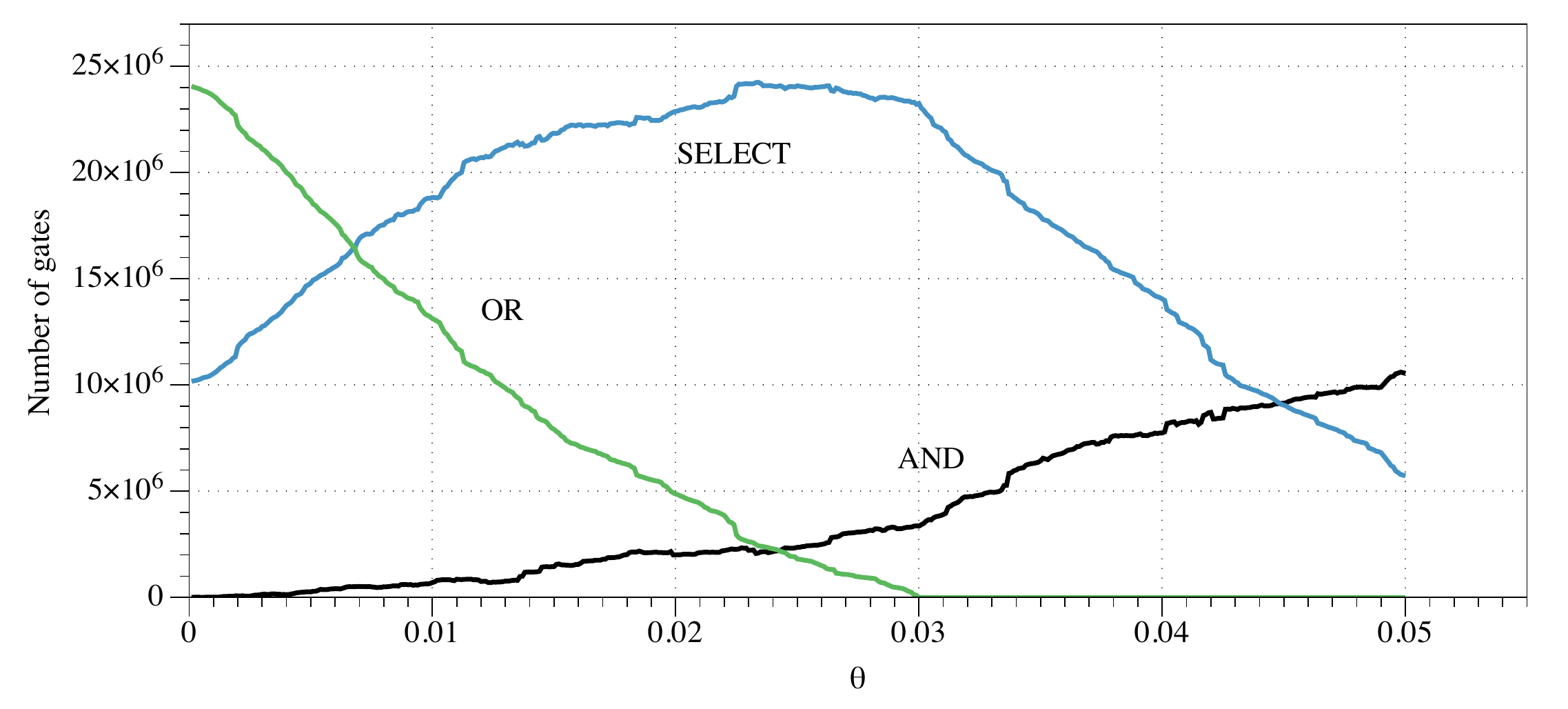}}
    \caption{Occurrences of the gates from the groups {\sc and}, black, {\sc or}, green, {\sc and-not}, red, and {\sc select}, blue, for $\theta \in [0.0001, 0.05]$, with $\theta$ increment 0.0001, in (a)~fungal colony modelled with serial RC networks, (b)~fungal colony modelled with parallel RC networks.}
    \label{fig:gates}
\end{figure}

In general, there are 16 possible logical gates realisable for two inputs and one output. The gates implying input $0$ and evoking a response $1$, i.e. $f(0,0)=1$, are not realisable because the fungal circuit simulated is passive. The remaining 8 gates are   
{\sc and}, {\sc or}, 
{\sc and-not} ($x$ {\sc and not} $y$ and {\sc not} $x$ {\sc and} $y$),
{\sc select}  ({\sc select} $x$ and {\sc select} $y$) 
and {\sc xor}.

No {\sc xor} gates have been found in neither of the RC models of the fungal colony.  

In the model of serial RC networks we found gates {\sc and}, {\sc select} and {\sc and-not}; no {\sc or} gates have been found. The number $n$ of the gates discovered decreases by a power low with increase of $\theta$: 
$n_{{\text {\sc and-not}}}=72 \cdot  x^{-0.98}$, $n_{{\text {\sc select}}}=2203 \cdot x^{-0.48}$, $n_{{\text {\sc and}}}=0.02 \cdot x^{-1.6}$. Frequency of {\sc and} gate oscillates, as shown in zoom insert in Fig.~\ref{fig:gates}a, more likely due to its insignificant presence in the samples. The oscillations reach near zero base when $\theta$ exceeds 0.001.

In the model of parallel RC networks we found only gates {\sc and}, {\sc select} and {\sc or}. The number of {\sc or} gates decreases quadratically and becomes nil when $\theta>0.03$. The number of {\sc and} gates increases near linearly, $n_{{\text {\sc and}}}= -1.72 \cdot 10^6 + 2.25 \cdot 10^8 \cdot x$, with increase of $\theta$. The number of {\sc select} gates reaches its maximum at $\theta=0.023$, and then starts to decreases with the further increase of $\theta$: $n_{{\text {\sc select}}}= 9.61 \cdot 10^6 + 1.21 \cdot 10^9 \cdot x - 2.7 \cdot x^2$. 

\section{Discussion}
\label{discussion}

By simulating a fungal colony as an electrical network we discovered families of Boolean gates realisable in the network. Voltage values have been binarised via threshold $\theta$.  All non-active, i.e. $f(0,0) \neq 1$, gates but {\sc xor} have been discovered and their dynamics in relation to $\theta$. The systems of gates discovered are functionally complete and therefore we can speculate that an arbitrary logical circuit can be realised in living fungal networks by encoding Boolean values in differences of electrical potential. The {\sc xor} gates have not been observed in our models. This is unsurprising as the {\sc xor} gate is the most rare gate to be discovered in non-linear systems~\cite{adamatzky2009complex,siccardi2016boolean,harding2018discovering}. A disadvantage of the electrical analog logical circuits in living fungal colonies would be that the colony requires  maintenance and have a relatively short life span. A way forward would be to `imprint' the colony in other long-living materials. This can be done, for example, by means of biolithography as previously tested on slime mould \emph{Physarum polycephalum}~\cite{berzina2018biolithography}.

\section{Acknowledgement}

This project has received funding from the European Union's Horizon 2020 research and innovation programme FET OPEN ``Challenging current thinking'' under grant agreement No 858132.

\section{Appendix}
\label{appendix}

\subsection{Fungal colony imaging}

\emph{Aspergillus niger} strain AR9\#2~\cite{vinck2011heterogenic}, expressing Green Fluorescent Protein (GFP) from the glucoamylase (\emph{glaA}) promoter, was grown at 30\textsuperscript{o}C on minimal medium (MM)~\cite{de2004new} with 25~mM xylose and 1.5\% agarose (MMXA). MMXA cultures were grown for three days, after which conidia were harvested using saline-Tween (0.8\% NaCl and 0.005\% Tween-80). 250~ml liquid cultures were inoculated with $1.25\cdot 10^9$ freshly harvested conidia and grown at 200~rpm and 30\textsuperscript{o}C in 1~L Erlenmeyer flasks in complete medium (CM) (MM containing 0.5\% yeast extract and 0.2\% enzymatically hydrolyzed casein) supplemented with 25~mM xylose (repressing glaA expression). Mycelium was harvested after 16 h and washed twice with PBS. Ten g of biomass (wet weight)  was transferred to MM supplemented with 25~mM maltose (inducing glaA expression). 

Fluorescence of GFP was localised in micro-colonies using a DMI 6000 CS AFC confocal microscope (Leica, Mannheim, Germany). Micro-colonies were fixed overnight at 4\textsuperscript{o}C in 4\% paraformaldehyde in PBS, washed twice with PBS and taken up in 50~ml PBS supplemented with 150~mM glycine to quench autofluorescence. Micro-colonies were then transferred to a glass bottom dish (Cellview\texttrademark, Greiner Bio-One, Frickenhausen, Germany, PS, 35/10 MM) and embedded in 1\% low melting point agarose at 45\textsuperscript{o}C. Micro-colonies were imaged at 20$\times$ magnification (HC PL FLUOTAR L 20 $\times$ 0.40 DRY). GFP was excited by white light laser at 472~nm using 50\% laser intensity (0.1~kW/cm2) and a pixel dwell time of 72~ns. Fluorescent light emission was detected with hybrid detectors in the range of 490–525~nm. Pinhole size was 1 Airy unit.

\subsection{Graph extraction}

3D projections were made with Fiji~\cite{schindelin2012fiji}. Conversion of the imaged micro-colonies to graph data was accomplished using a publicly available ImageJ macro\footnote{The macro was developed by the Advanced Digital Microscopy Core Facility, IRB Barcelona, to process Z-stack data for blood vessel segmentation and network analysis, see details in \url{adm.irbbarcelona.org/bioimage-analysis/image-j-fiji} and \url{biii.eu/blood-vessel-segmentation-and-network-analysis}}.
The macro was run on the Fiji (ImageJ) platform, version 1.52, with the supplementary 3D ImageJ Suite installed to provide enhanced 3D capabilities \cite{ollion2013tango}.  To run the macro, initialisation parameters for expected hypha radius were given as 3~µm \cite{tegelaar2017functional}; detection sensitivity threshold was set to 8; minimal vessel volume was set to 100 pixels. Key processing tasks performed by the macro were:   \begin{enumerate}
    \item Morphological closing of tubular structures.
    \item Pre-filtering to enhance filamentous voxels.
    \item Segmentation of tubular structures.
    \item Skeletonisation and analysis of the network. 
\end{enumerate}
Results of the segmentation included a Z-stack 3D visualisation of the network with identified branching and end-points, and tabular data including the number of disjoint networks together with constituent branch segments defined by start and end points given in voxel coordinates.

The tabular data was then processed using a custom python script to convert voxel coordinates to real-world coordinates. The set of all vertices was determined and all branch start and end points were indexed. A weighted graph was generated using the NetworkX library for python \cite{hagberg2008exploring}, with graph nodes defined by the vertex index, edges defined between vertex pairs and edge weights given as the euclidean distance of the branch segment (Fig.~\ref{fig:graph}). 
Graph topology could then be determined using the NetworkX degree function. Shortest weighted paths between source and sink vertices could also be found, allowing a direct correlation to resistive networks.


\end{document}